\documentclass[secnumarabic, graphics,floatfix, nofootinbib,tightenlines,nobibnotes, aps, prl, 12pt]{revtex4-2}
\usepackage[english]{babel}
\usepackage[utf8]{inputenc}
\usepackage{amsfonts}
\usepackage{multirow}
\usepackage{graphicx}
\usepackage{epstopdf}
\usepackage{hyperref}
\usepackage{latexsym}
\usepackage{amsmath}
\usepackage{amssymb}

\usepackage[utf8]{inputenc}
\hypersetup{
    colorlinks=true,
    linkcolor=red,
    citecolor=blue,
}
\usepackage{color}
\usepackage[T1]{fontenc}
\usepackage{txfonts}
\usepackage{mathrsfs}
\usepackage[center]{subfigure}

\begin{document}
 \setcounter{secnumdepth}{2}
 \newcommand{\bq}{\begin{equation}}
 \newcommand{\eq}{\end{equation}}
 \newcommand{\bqn}{\begin{eqnarray}}
 \newcommand{\eqn}{\end{eqnarray}}
 \newcommand{\nb}{\nonumber}
 \newcommand{\lb}{\label}
 
\title{Matrix method for perturbed black hole metric with discontinuity}

\author{Shui-Fa Shen$^{1,2}$}\email[E-mail: ]{shuifa.shen@inest.cas.cn}
\author{Wei-Liang Qian$^{3,4,5}$}\email[E-mail: ]{wlqian@usp.br (corresponding author)}
\author{Kai Lin$^{6,3}$}\email[E-mail: ]{lk314159@hotmail.com}
\author{Cheng-Gang Shao$^{7}$}\email[E-mail: ]{cgshao@hust.edu.cn}
\author{Yu Pan$^{8}$}\email[E-mail: ]{panyu@cqupt.edu.cn}

\affiliation{$^{1}$ School of intelligent manufacturing, Zhejiang Guangsha Vocational and Technical University of Construction, 322100, Jinhua, Zhejiang, China}
\affiliation{$^{2}$ School of Electronic, Electrical Engineering and Physics, Fujian University of Technology, 350118, Fuzhou, Fujian, China}
\affiliation{$^{3}$ Escola de Engenharia de Lorena, Universidade de S\~ao Paulo, 12602-810, Lorena, SP, Brazil}
\affiliation{$^{4}$ Faculdade de Engenharia de Guaratinguet\'a, Universidade Estadual Paulista, 12516-410, Guaratinguet\'a, SP, Brazil}
\affiliation{$^{5}$ Center for Gravitation and Cosmology, School of Physical Science and Technology, Yangzhou University, 225002, Yangzhou, Jiangsu, China}
\affiliation{$^{6}$ Hubei Subsurface Multi-scale Imaging Key Laboratory, Institute of Geophysics and Geomatics, China University of Geosciences, 430074, Wuhan, Hubei, China}
\affiliation{$^{7}$ MOE Key Laboratory of Fundamental Physical Quantities Measurement, Hubei Key Laboratory of Gravitation and Quantum Physics, PGMF, and School of Physics, Huazhong University of Science and Technology, 430074, Wuhan, Hubei, China}
\affiliation{$^{8}$ College of Science, Chongqing University of Posts and Telecommunications, 400065, Chongqing, China}

\date{June 20th, 2022}

\begin{abstract}
Recent studies based on the notion of black hole pseudospectrum indicated substantial instability of the fundamental and high-overtone quasinormal modes.
Besides its theoretical novelty, the details about the migration of the quasinormal mode spectrum due to specific perturbations may furnish valuable information on the properties of associated gravitational waves in a more realistic context.
This work generalizes the matrix method for black hole quasinormal modes to cope with a specific class of perturbations to the metric featured by discontinuity, which is known to be intimately connected with the quasinormal mode structural instability.
In practice, the presence of discontinuity poses a difficulty so that many well-known approaches for quasinormal modes cannot be straightforwardly applied.
By comparing with other methods, we show that the modified matrix method is efficient, which can be used to solve for the low-lying modes with reasonable precision. 
Therefore, it might serve as an alternative gadget for relevant studies.

\end{abstract}

\maketitle

\newpage

\section{Introduction}\label{section1}

As the culmination of continuous efforts during the past few decades, the advent of empirical detection of the gravitational wave inaugurated a novel era of gravitational wave (GW) astronomy.
The GWs provide unique information on the dynamics of the source that is complementary to those obtained via electromagnetic radiation. 
In particular, the merger signals captured by the LIGO and Virgo collaboration~\cite{agr-LIGO-01, agr-LIGO-02, agr-LIGO-03, agr-LIGO-04} furnished direct evidence of black holes and has promoted further interest in the ongoing space-borne missions,
such as LISA~\cite{agr-LISA-01}, TianQin~\cite{agr-TianQin-01}, and Taiji~\cite{agr-Taiji-01}.
As a result, significant efforts have been devoted to modeling systems composed of compact astrophysical objects, such as binaries of black holes or neutron stars, as possible sources of gravitational radiations.
In terms of dissipative oscillations, black hole quasinormal modes~\cite{agr-qnm-review-02, agr-qnm-review-03, agr-qnm-review-06} (QNMs) primarily constitute the ringdown phase of a merger event.
These temporal profiles carry the essential spacetime properties of the resulting black hole metric, subject to a few no-hair theorems~\cite{agr-bh-nohair-01, agr-bh-nohair-04}.
In its own right, topics regarding QNM continue to attract much attention in the recent years~\cite{Santos:2019yzk, Campos:2021sff, Anacleto:2021qoe, Rincon:2018sgd, Panotopoulos:2017hns, Rincon:2018ktz, Ponglertsakul:2018smo, Fernando:2022wlm, Gonzalez:2022upu, Rincon:2021gwd}.

GW astronomy as a precision discipline might potentially be interfered with by a few crucial factors.
On the detector side, for the GW frequency range aimed by the space-borne interferometers (about $0.1\rm{mHz-kHz}$), the primary noise sources are comprised of laser frequency fluctuations, mechanical vibrations, and clock jitters~\cite{agr-TDI-review-01, agr-TDI-review-02, agr-TDI-Wang-01}.
Appropriate noise suppression schemes are under active development.
In practice, the resultant sensitivity can be formulated in terms of the signal-to-noise ratio~\cite{agr-SNR-Wang-01, agr-SNR-Wang-02}, and pertinent studies have been performed concerning the feasibility of an eventual detection~\cite{agr-SNR-05, agr-SNR-06, agr-SNR-10, agr-SNR-18, agr-SNR-20, agr-SNR-36}.
On the other hand, a black hole or neutron star is not an isolated object from the astrophysical viewpoint.
To be specific, in a realistic environment, they are exposed to various surrounding matters such as the accretion disk, electromagnetic field, quintessence fluid, or cosmological expansion.
The resulting metric and, subsequently, the QNMs, are expected to deviate from those of an isolated compact object, which subsequently leads to the notion of ``dirty'' black holes~\cite{agr-bh-thermodynamics-12, agr-qnm-33, agr-qnm-34, agr-qnm-54}.

In~\cite{agr-qnm-33, agr-qnm-34}, Leung {\it et al.} investigated the scalar QNMs in perturbated nonrotating black holes.
The deviations in both the real and imaginary parts of the quasinormal frequencies from those of an isolated black hole are obtained using the generalized logarithmic perturbation theory.
Barausse {\it et al.} carried out a detailed analysis of the ringdown and inspiral processes due to small perturbation about a central Schwarzschild black hole~\cite{agr-qnm-54}.
Regarding QNMs, the authors observed that the resultant QNMs might differ substantially from those of the isolated black hole.
However, even though the latter does not show up in the spectrum, as poles of the relevant Green's function, of the dirty black hole metric, these modes still play an essential role in the time profile.
Moreover, the authors concluded that the astrophysical environment does not significantly impact the feasibility of GW spectroscopy by using templates of isolated black holes.
It is worth noting that among the various scenarios explored in~\cite{agr-qnm-54}, the thin shell model turns out to have the most prominent modification to the QNM spectrum.

From a somewhat different but intriguing perspective, earlier, Nollert~\cite{agr-qnm-35} investigated the effect of a series of small-scale perturbations on the QNM spectrum. 
In particular, the entire Regge-Wheeler potential was approximated by a series of step functions, and the resulting temporal evolutions and QNMs were examined.
An overall instability of the QNM spectrum was observed, which was then interpreted as an apparent contradiction by the author.
Such a conclusion resides on the understanding that once a reasonably accurate approximation is adopted for the effective potential, the resulting physics is not expected to be drastically different.
Following this line of thought, the problem was further explored by Nollert and Price~\cite{agr-qnm-36} for spiked truncated dipole potential, 
Daghigh {\it et al.}~\cite{agr-qnm-50} introduced a continuous piecewise linear potential to improve the approximation to the Regge-Wheeler potential, but the main features persist in these models.
In~\cite{agr-qnm-lq-03}, some of us pointed out that as long as discontinuity is present in the effective potential, the asymptotic high-overtone QNMs will always tend to stretch out along the real axis.
As shown analytically, the result holds even when the discontinuity is located significantly further away from the horizon, or the step size is arbitrarily insignificant. 
This implies that the asymptotic properties of the high-overtone modes are intrinsically different from those of the isolated black hole since their counterparts mostly line up parallel to the imaginary axis~\cite{agr-qnm-continued-fraction-02, agr-qnm-continued-fraction-03}.
Furthermore, Jaramillo {\it et. al}~\cite{agr-qnm-instability-07, agr-qnm-instability-13, agr-qnm-instability-14} analyzed the problem by employing the notion of structural stability.
Specifically, the stability of the quasinormal mode spectrum was investigated under randomized perturbations to the metric.
The numerical calculations were carried out using Chebyshev's spectral method in the hyperboloidal coordinates~\cite{agr-qnm-hyperboloidal-01}.
It was observed that the boundary of the pseudospectrum triggered by {\it ultraviolet} perturbations migrates towards the real frequency axis.
As a more systematic approach, the obtained results agree with the existing studies while indicating a universal instability for the high-overtone modes.
More recently, it was demonstrated by Cheung {\it et al.}~\cite{agr-qnm-instability-15} that even the fundamental mode can be destabilized under generic perturbations to the Schwarzschild potential.
The phase diagrams of a simple double-barrier toy model were analyzed to clarify the specific conditions under which the instability takes place.
In light of the above results, any conclusion drawn from the analysis solely based on the QNM of the isolated black hole might largely be undermined if an insignificant change in the system will cause a drastic modification to the QNM spectrum.
In other words, unlike the more conventional approaches in black hole perturbation theory, where one explores the stability of the underlying metric through the quasinormal modes, these recent studies indicate the significance regarding the stability of the quasinormal mode spectrum itself.
However, as pointed out in~\cite{agr-qnm-instability-07}, for the results to be meaningful in a realistic context, the relevant perturbations to the non-selfadjoint operator in question must be physically plausible rather than arbitrary ones.
Therefore, in order to ascertain whether the system is beset with spectral instability, one has to carefully study the migration of the quasinormal spectrum under perturbations corresponding to those in a realistic environment.
A possible strategy is to further explore pertinent scenarios of metric perturbations, whose impacts are potentially significant.
Among others, as a mathematically simple and physically relevant feature, discontinuity as a particular type of perturbations to the black hole's effective potential is a worthy topic.

Notably, many conventional approaches for the quasinormal modes cannot be straightforwardly employed to deal with discontinuity.
For instance, the standard WKB formulae~\cite{agr-qnm-WKB-01, agr-qnm-WKB-02, agr-qnm-WKB-03, agr-qnm-WKB-05} evaluate the quasinormal frequencies based only on the information of the effective potential (inclusively its derivatives) at its maximum.
Also, the monodromy method~\cite{agr-qnm-40} is based on the analytic continuation of the wave function in coordinate space.
Thus discontinuity introduces complications to the analytic properties of the waveform.
Moreover, the assumption $\Im \omega \gg \Re \omega$ also becomes invalid for asymptotic modes with large real parts, relevant for unstable spectrum.
In the study of QNMs in pulsating relativistic stars for which discontinuity occurs at the surface, Kokkotas and Schutz~\cite{agr-qnm-star-07} utilized the numerical integration, while Leins {\it el al.}~\cite{agr-qnm-star-08} modifies Leaver's continued fraction method~\cite{agr-qnm-continued-fraction-01}. 
The latter has been inherited in a few subsequential studies~\cite{agr-qnm-34, agr-qnm-star-25, agr-qnm-54}.
It turned out to be a dependable method and can be used to handle discontinuous effective potential.

The matrix method~\cite{agr-qnm-lq-matrix-01,agr-qnm-lq-matrix-02,agr-qnm-lq-matrix-03,agr-qnm-lq-matrix-04} proposed by some of us is an approach that reformulates the QNM problem into a matrix equation for the complex frequencies.
To some extent, it is reminiscent of the continued fraction method, and their main difference resides in that the expansion of the wave function is carried out about a series of discrete coordinate grids.
However, the method cannot be applied directly to an effective potential that contains discontinuous points.
In this work, we generalize the matrix method for black hole quasinormal modes to deal with a specific class of metrics featured by discontinuity.
By comparing the results obtained by other approaches, we show that the modified matrix method is capable of solving for the low-lying modes efficiently with desirable precision.
Therefore, it might serve as an alternative tool for relevant studies.

The remainder of the paper is organized as follows.
The following section gives a brief account of the matrix method and reviews its main technical features.
The modifications to the formulae are discussed in Sec.~\ref{section3}.
The matrix is revised in accordance with the connection conditions at the boundaries in terms of the Wronskian.
In Sec.~\ref{section4} we discuss a few specific examples, which consist of a toy model of the square potential barrier and the thin shell models for Schwarzschild black hole metric.
The results are compared with the other approaches, including the modified continued fraction method primarily encountered in the literature.
Further discussions and the concluding remarks are given in Sec.~\ref{section5}.

\section{The matrix method}\label{section2}

The strategy of the matrix method~\cite{agr-qnm-lq-matrix-02,agr-qnm-lq-matrix-03} essentially follows Leaver's continued fraction approach~\cite{agr-qnm-continued-fraction-01}.
The primary difference resides in the fact~\cite{agr-qnm-lq-matrix-01} that, in place of the Taylor expansions of the wavefunction around the horizon, a series of expansions are carried out at discrete grid points.
Moreover, in principle, these grid points do not necessarily distribute evenly in the entire domain of the tortoise coordinate.
In other words, when such freedom is appropriately adjusted, one manages to increase the ``resolution'' of the wavefunction for the more important region.
In terms of the above expansions, one discretizes the master equation of the perturbation field and rewrites it in the form of a matrix equation.
As a non-linear algebraic equation, the latter can be solved for the complex quasinormal frequencies by various algorithms.
Another nice feature of the matrix method is that it can be readily applied to the case where the master equation is furnished by a system of coupled degrees of freedom~\cite{agr-qnm-lq-matrix-03}.

The matrix method handles the boundary conditions of the QNM problem in a similar fashion as the continued fraction method. 
First, one rewrites the master equation by subtracting the asymptotic form of the waveform, obtained by using the ingoing and outgoing boundary conditions, respectively, at the horizon and outer spatial bound (for instance, the spatial infinity).
However, unlike the continued fraction method that utilizes the convergent criterion regarding the recurrence relations for the coefficients, the expansions of the matrix method are truncated.
Similar to other methods based on Taylor expansion, one transforms the domain of the tortoise coordinate $r_*$ into a finite range $z\in[0,1]$, where the boundaries correspond to the endpoints $z=0$ and $z=1$, respectively.
To be specific, by introducing the above transforms, the standard form of the master equation~\cite{agr-qnm-review-02}
\begin{equation}\lb{QNMaster}
\left[\frac{\partial^2}{\partial r_*^2}+\omega^2-V_{\mathrm{eff}}\right]\Psi = 0 ~,
\end{equation}
where $V_{\mathrm{eff}}(z)$ is effective potential, can be written\footnote{For an explicit example, see Eqs.~\eqref{zV2},~\eqref{psiV2}, and~\eqref{RV2} below.} as~\cite{agr-qnm-lq-matrix-04}
\bqn
\lb{masterMM}
{\cal H}(\omega, z)R(z)=0 ~,
\eqn
with the boundary conditions
\bqn
\lb{4}
R(z=0)=C_0~~~\text{and}~~~R(z=1)=C_1 ~.
\eqn
Here ${\cal H}(\omega, z)$ is a linear operator, that is essentially governed by the effective potential $V_{\mathrm{eff}}$ of the original master equation.
$C_0$ and $C_1$ are constants since the asymptotic part of the wave function has already been subtracted.
For convenience, one further introduces
\bqn
\lb{5}
F(z)=R(z){z(1-z)} ~,
\eqn
and rewrites Eq.~\eqref{masterMM} as
\bqn
\lb{qnmeqf}
{\cal G}(\omega,z)F(z)=0 ~,
\eqn
with simpler boundary conditions
\bqn
\lb{qnmbcf}
F(z=0)=F(z=1)=0 ~.
\eqn
We note that Eq.~\eqref{5} introduces additional irrelevant roots, which usually do not pose a problem in practice, as long as they stay away from the quasinormal frequencies.

To proceed, one descretizes Eq.~(\ref{qnmeqf}) by introducing $N$ grid points $z_i$ (with $i=1, 2, \cdots, N$), at which the expansions are performed. 
The values of the wave function $F(z)$ on the grids are denoted by $f_i=F(z_i)$.
Now, the crucial step is to discretize the operator ${\cal G}(\omega,z)$ and reformulate Eq.~(\ref{qnmeqf}) into a matrix equation.
This can be accomplished by repeatedly employing the expansion on individual grids $z_i$ and reverting the resultant $N\times N$ matrix as discussed in~\cite{agr-qnm-lq-matrix-01}.
We note that the above process is universal as long as the number of total grids is given. 
The obtained matrix equation can be formally written as
\bqn
\lb{7}
\overline{\cal M}\ {\cal F}=0 ~,
\eqn
where $\overline{\cal M}$ represents an $N\times N$ matrix which still is a function of the frequency $\omega$, and
\bqn
{\cal F}=(f_1,f_2,\cdots,f_i,\cdots,f_N)^T
\eqn
is a column vector composed of $f_i$.

To accomodate the boundary conditions Eq.~(\ref{qnmbcf}), which implies that
\bqn
\lb{8}
f_1=f_N=0~,
\eqn
one {\it amends} Eq.~(\ref{7}) to read
\bqn
\lb{qnmeqMatrix}
{\cal M}\ {\cal F}=0~,
\eqn
where the element ${\cal M}_{k,i}$ of the matrix ${\cal M}$ is defined by
\bq
\lb{10}
{\cal M}_{k,i}= 
\left\{\begin{array}{cc}
\delta_{k,i},     &  k=1~\text{or}~N \cr\\
\overline{\cal M}_{k,i}, &  k=2,3,\cdots,N-1
\end{array}\right. ~.
\eq
The obtained matrix equation indicates that ${\cal F}$ is the eigenvector of ${\cal M}$, which gives
\bqn
\lb{qnmDet}
\det {\cal M}(\omega) = 0~.
\eqn
In general, Eq.~(\ref{qnmDet}) is a non-linear algebraic equation for the quasinormal frequency $\omega$, which can be handled by the standard algebraic equation solver.

\section{The modifications to the matrix due to discontinous boundaries} \lb{section3}

Now, we turn to discuss the generalization of the above algorithm for the effective potential that contains discontinuity.
Let us consider that at one of the grid points $z = z_c$\footnote{If the discontinuity does not occur at a grid point, one can always rearrange the grid points so that one of them sits on the discontinuity.}, the effective potential is discontinuous.
Such a discontinuity can be viewed as a limit of some realistic scenario governed by Israel's junction condition~\cite{agr-collapse-thin-shell-03, book-general-relativity-Poisson}.
This leads to two immediate consequences.
First, the Taylor expansion becomes invalid in the entire domain.
Second, the wave functions on the two sides of discontinuity are related by~\cite{agr-qnm-34}
\bqn
\lb{Israel}
\lim_{\epsilon\to 0^+}\left[\frac{R'(z_c+\epsilon)}{R(z_c+\epsilon)}-\frac{R'(z_c-\epsilon)}{R(z_c-\epsilon)}\right]=\kappa \ ,
\eqn
where, for the Schr\"odinger-type master equation Eq.~\eqref{QNMaster},
\bqn
\lb{kappa}
\kappa = \lim_{\epsilon\to 0^+}\int_{z_c-\epsilon}^{z_c+\epsilon} V_{\mathrm{eff}}(z)dz \ .
\eqn
If one considers a moderate finite jump, the above relation simplifies to the condition of vanishing Wronskian
\bqn
\lb{Wronskian}
W(\omega) \equiv R(z_c+\epsilon)R'(z_c-\epsilon) - R(z_c-\epsilon)R'(z_c+\epsilon) = 0~.
\eqn

The above elements can be adequately taken into account by revising the matrix ${\cal M}$ in Eq.~\eqref{10} as follows.
Since the Taylor expansion is no longer valid for the entire region, they must only be carried out for individual intervals of $z$ where the potential is continuous.
Subsequently, this corresponds to that the matrix ${\cal M}$ is {\it almost} broken into diagonal sections of block submatrices.
Moreover, the relation Eq.~\eqref{Israel} or~\eqref{Wronskian} can be implemented on the line between two successive blocks.
Specifically, one replaces the information on the waveforms of respective regions and their first-order derivatives with the discrete values of the entire wave function on the grids.
In practice, if the above boundary corresponds to the $i$th grid, then the $i$th line and column of the matrix ${\cal M}$ are occupied by both blocks.
On the one hand, the $i$th line corresponds to an equation governed by the relation between waveforms on the boundary.
On the other, for the case of Eq.~\eqref{Wronskian}, the $i$th column is filled with elements of both blocks because the value of the waveform on the $i$th grid readily enters the interpolation schemes in both regions.
Therefore, the modified matrix ${\cal M}$ is not entirely block diagonalized.
This result can also be understood because, otherwise, the resulting QNM spectrum would be constituted by a simple summation of those pertaining to individual blocks.

\section{Numerical results}\lb{section4}

This section presents a few numerical examples using the modified matrix method described in the last section.
To be specific, we discuss three specific forms for the effective potential $V_{\mathrm{eff}}$ that are featured by discontinuity and evaluate the QNM of the master equation Eq.~\eqref{QNMaster}.
The obtained results are then compared with those obtained by other approaches.

\begin{figure*}[htbp]
\centering
\includegraphics[scale=0.8]{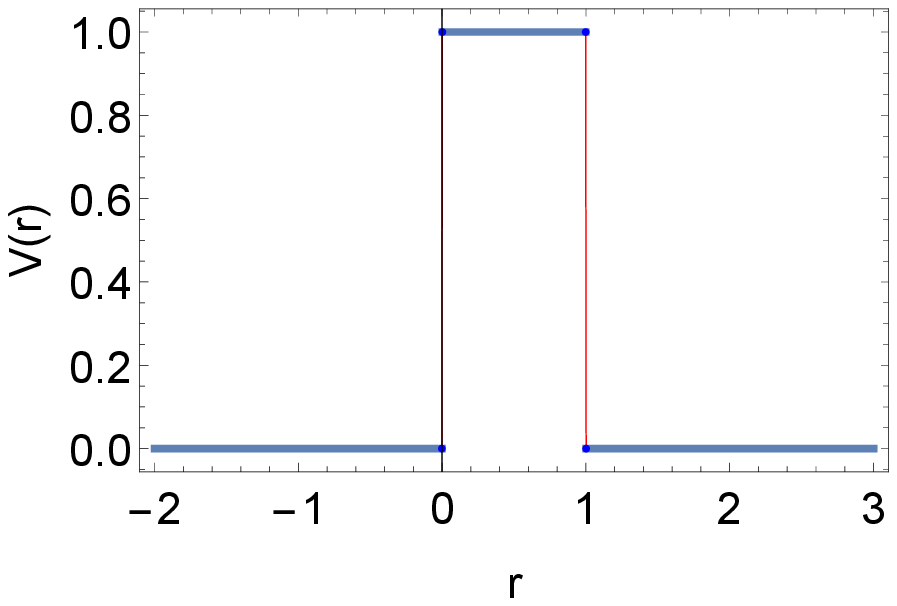} \ \ \ \ 
\includegraphics[scale=0.8]{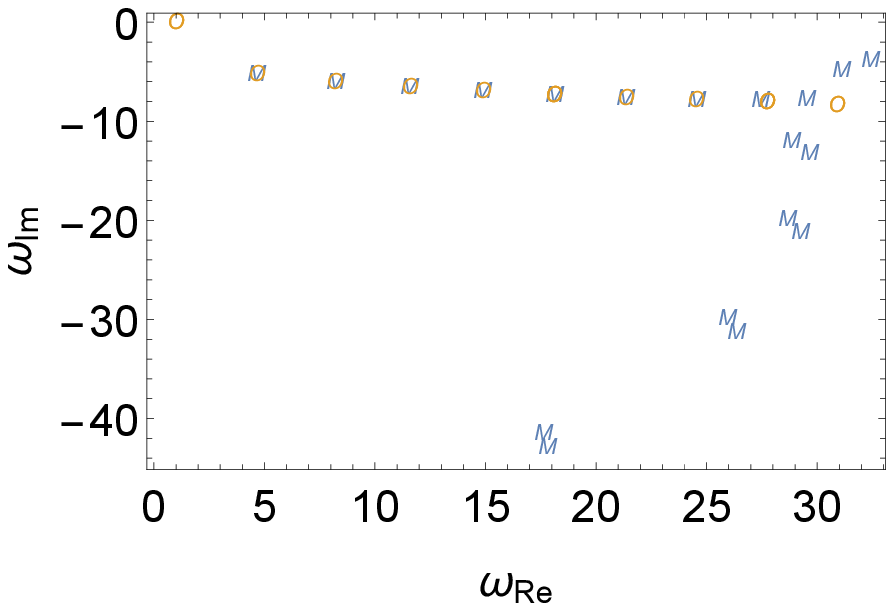}
\caption{
Left: The effective potential of a simple square barrier.
Right: The obtained QNMs using the matrix method (indicated by the blue ``$M$''s) compared against those obtained by a straightforward root-finding scheme (indicated by the orange ``$O$''s).
}
\label{Fig1}
\end{figure*}

\begin{table}[]
\caption{The low-lying QNM extracted for the case of square potential barrier using the generalized matrix method (MM) and a more precise root-finding (RF) scheme.}\lb{Tab1}
\begin{ruledtabular}
\renewcommand\arraystretch{2}
\begin{tabular}{cccccccccc}
                        & $n$			& $0$		& $1$	& $2$	           & $3$ 	& $4$    		& $5$		& $6$		     & $7$ \\
\hline
MM  & $\omega_{\mathrm{Re}}$ & $4.63002$   & $8.19758$  & $11.5675$ & $14.857$  &  $18.104$    & $21.3248$  &  $24.5276$  & $27.4459$ \\
                           & $\omega_{\mathrm{Im}}$ & $-5.21893 i$  & $-5.99133 i$ & $-6.54004 i$ & $-6.96791 i$ & $-7.3194 i$   & $-7.61804 i$ & $-7.88782 i$ & $-7.86098 i$ \\
\hline
RF & $\omega_{\mathrm{Re}}$ & $4.630020$  & $8.1975809$  & $11.56754$       & $14.8570$  & $18.10397$ & $21.325125$  & $24.52928$  & $27.72158$ \\
                          & $\omega_{\mathrm{Im}}$ & $-5.218929 i$ & $-5.991330 i$ & $-6.540041 i$ & $-6.967907 i$ & $-7.319397 i$ & $-7.6179780 i$ & $-7.8776369 i$ & $-8.107418 i$  
\end{tabular}
\end{ruledtabular}
\end{table} 

For the first numerical example, we evaluate the low-lying QNM for a toy model of square potential barrier 
\bqn
\lb{Veff1}
V_{\mathrm{eff}} = V_1(r)=\left\{ \begin{matrix}
0&r_*<0\\
1&0\le r_*\le 1\\
0&r_*>1
\end{matrix} \right. \ ,
\eqn
where $r_*=r$, as shown in the left plot of Fig.~\ref{Fig1}.
To employ the matrix method, one utlizes 31 grid points.
For the regions $z=r_*>1$ and $z=r_*<0$, the waveforms are the right-(out)going and left-(in)going plane waves, $R_1=e^{i\omega r_*}$ and $R_2=e^{-i\omega r_*}$, respectively. 
By Eq.~\eqref{Wronskian}, the ratios
\bqn
\lb{Veff1bd}
\frac{R'}{R} = \left\{ \begin{matrix}
-i\omega &z=0\\
&\\
i\omega&z=1
\end{matrix} \right.
\eqn
can be used to replace the first and last lines of the matrix ${\cal M}$ by appropriately expressed in terms of the values of the wave function on the grids.
Subsequently, the determinant Eq.~\eqref{qnmDet} furnishes the relevant quasinormal frequencies.

On the other hand, the QNMs of the square potential barrier can also be calculated by a straightforward approach.
Owing to the simple form of the effective potential, one can analytically evaluate the transmission matrix $\mathbb{T}$ (usually defined from the l.h.s. to the r.h.s.) of the potential Eq.~\eqref{Veff1}.
The resulting QNMs correspond to the roots of the numerator of $\mathbb{T}_{22}=0$~\cite{agr-qnm-Poschl-Teller-02}, namely,
\bqn
\lb{Veff1root}
e^{2i \sqrt{-1+\omega^2}}\left[1+2\omega\left(-\omega+\sqrt{-1+\omega^2}\right)\right]+2\omega \left(\omega+\sqrt{-1+\omega^2}\right)-1 = 0~.
\eqn
In practice, to seek the roots of Eq.~\eqref{Veff1root}, the complex frequencies obtained by the matrix method are fed to an equation solver as the initial values.

\begin{figure*}[htbp]
\centering
\includegraphics[scale=0.8]{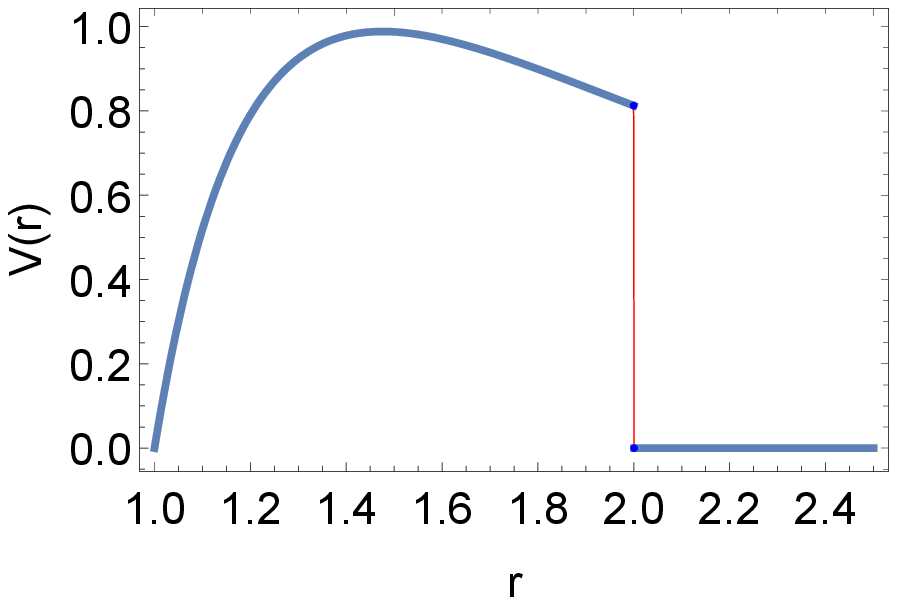} \ \ \ \ 
\includegraphics[scale=0.8]{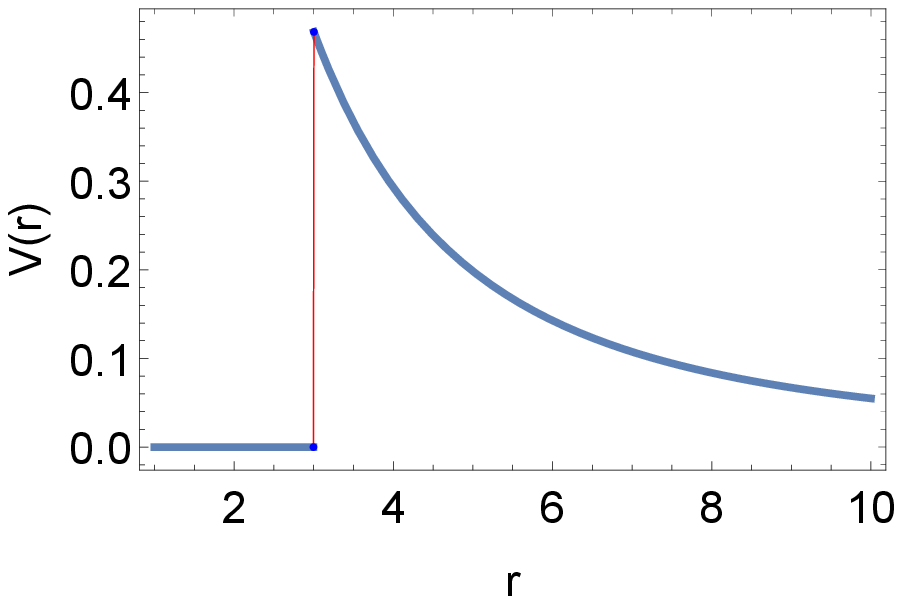}
\caption{
Left: The effective potential for axial gravitational perturbations in Schwarzschild black hole metric where a cut is implemented at $r_c=2$ for the outer region.
Right: A similarly devised effective potentail with a cut at $r_c=3$ but for the inner region.}
\label{Fig2}
\end{figure*}

In Tab.~\ref{Tab1} and Fig.~\ref{Fig1}, we show the obtained low-lying QNMs, which are compared against those obtained by solving Eq.~\eqref{Veff1root}.
It is observed that the low-lying QNMs are reasonably reproduced.
It is noticeable from Fig.~\ref{Fig2} that the root-finding scheme also encounters the root $\omega = 1$, which must be excluded because it is physically irrelevant for its appearance as a root in the denominator of $\mathbb{T}_{22}$.
Moreover, this simple case seems to indicate the asymptotic behavior of the high-overtone modes observed in the literature.

The effective potential of the second example is devised by truncating the tail of the Regge-Wheeler potential.
To be specific, as shown in the left plot of Fig.~\ref{Fig2}, one defines
\bqn
\lb{Veff2}
V_{\mathrm{eff}} = V_2(r)=\left\{ \begin{matrix}
V_{\mathrm{RW}}& r\le r_c\\
0&r>r_c
\end{matrix} \right. \ ,
\eqn
where
\begin{equation}\label{eqRW}
V_{\mathrm{RW}}(r)=\left(1-\frac{r_h}{r}\right)\left[\frac{\ell(\ell+1)}{r^2}+(1-s^2)\frac{r_h}{r^3}\right] .
\end{equation}
For simplicity, we assume the parameters $r_h = 2M = 1$, $s=2$, and $\ell=2$ for axial gravitational perturbations.

To employ the matrix method, 31 grid points have been used.
We adopt the coordinate
\begin{equation}\label{zV2}
z = \frac{1-\frac{r_h}{r}}{1-\frac{r_h}{r_c}} ,
\end{equation}
so that the cut $r=r_c$ is placed at $z=1$.
As in~\cite{agr-qnm-continued-fraction-01}, the wave function $\Psi$ in Eq.~\eqref{QNMaster} is transformed by factoring out its asymptotic forms
\begin{equation}\label{psiV2}
\Psi = (r-r_h)^{-i\omega r_h}e^{i\omega(r-r_h)}r^{2i\omega r_h}\Phi 
\end{equation}
and
\begin{equation}\label{RV2}
R=z \Phi ~. 
\end{equation}
The resulting master equation of $R$ as a function of $z$ is discretized into the matrix form Eq.~\eqref{masterMM}.

The boundary condition at $r=r_c$ can be determined by the outgoing wave in the tortoise coordinate
\begin{equation}\label{goutV2}
g(r_*) = e^{i\omega r_*}~, 
\end{equation}
where
\begin{equation}\label{rstar}
r_* = r + r_h \ln (r-r_h)~. 
\end{equation}
Therefore one has
\bqn
\lb{Veff2bd}
\frac{R'}{R} = 1+2i r_h \omega  ~,
\eqn
where the derivative is taken with respect to $z$ at the point $z=1$.
Again, Eq.~\eqref{Veff2bd} is utlized to replace the last line of the matrix ${\cal M}$, and the zeros of the determinant Eq.~\eqref{qnmDet} gives rise to the quasinormal frequencies.

To confirm the obtained QNMs, reminiscent of the continued fraction method, one may carry out a Taylor expansion of the wave function in $z$ around the horizon $z=0$. 
The expansion coefficients subsequently satisfy a recurrence relation.
However, for the present case, it seems not straightforward to obtain a recurrence relation among a given number of terms that further simplifies to a three-term relation.
Nonetheless, the expansion coefficients can be obtained recursively and used to solve for the quasinormal modes~\cite{agr-qnm-lq-07}.

\begin{table}[]
\caption{The low-lying QNM extracted for the truncated Regge-Wheeler potential Eq.~\eqref{Veff2} obtained by using the generalized matrix method (MM) and a recurence Taylor expansion (RT) scheme.}\lb{Tab2}
\begin{ruledtabular}
\renewcommand\arraystretch{2}
\begin{tabular}{ccccccccccc}
                        & $r_c$			& $2.0$		& 	& 	           & $1.8$ 	&     		& 		& $1.5$		     &  & \\
                        & $n$			& $0$		& $1$	& $2$	           & $0$ 	& $1$    		& $2$		& $0$		     & $1$		&$2$ \\
\hline
MM  	           & $\omega_{\mathrm{Re}}$ & $0.725951$   & $0.814147 $  & $1.03538 $ &  $0.664385 $  & $0.676746 $ & $0.81442 $ & $0.505596$  &  $0.339348$    & $0.262501 $  \\
                           & $\omega_{\mathrm{Im}}$ & $-0.35518 i$  & $-1.10893 i$ & $-1.93394 i$ & $-0.388176 i$ & $-1.20204 i$ & $-2.10075 i$ & $-0.421822 i$ & $-1.29596 i$   & $-2.26398 i$ \\
\hline
RT	          & $\omega_{\mathrm{Re}}$ & $0.725953$  & $0.814156$  & $1.03627$ & $0.664385 $  & $0.676745 $ & $0.814418$  & $0.505596$  & $0.339348$ & $0.262594$  \\
                          & $\omega_{\mathrm{Im}}$ & $-0.355202 i$ & $-1.10896 i$ & $-1.93401 i$ & $-0.388177 i$ & $-1.20204 i$  & $-2.10076 i$ & $-0.421822 i$ & $-1.29596 i$ & $-2.2641 i$ 
\end{tabular}
\end{ruledtabular}
\end{table} 

In Tab.~\ref{Tab2}, we show the obtained low-lying QNMs for different cuts $r_c$, which are compared against those obtained by the recurrence Taylor expansion scheme.
It is observed that the results of the low-lying modes are in good agreement.
Also, we note that the metric configuration of the second example has been explored in the literature regarding the high-overtone modes.
The properties of the asymptotic $w$-modes have been investigated analytically in~\cite{agr-qnm-lq-03}, which was understood to be related to the spectral instability~\cite{agr-qnm-instability-07}.

For the third example, we consider another type of modified Regge-Wheeler potential.
In this case, as shown in the right plot of Fig.~\ref{Fig2}, the potential is truncated from inside, to be specific, one defines
\bqn
\lb{Veff3}
V_{\mathrm{eff}} = V_3(r)=\left\{ \begin{matrix}
V_{\mathrm{RW}}& r\ge r_c\\
0&r<r_c
\end{matrix} \right. \ .
\eqn
Again, for simplicity, we assume the parameters $r_h = 2M = 1$, $s=2$, and $\ell=2$ for axial gravitational perturbations.

The matrix method can be employed in a very similar fashion.
We adopt 31 grid points for the numerical calculations, and the coordinate is defined as
\begin{equation}\label{zV3}
z = 1-\frac{r_c}{r} ,
\end{equation}
so that the cut $r=r_c$ is placed at $z=0$.
Similarly, the wave function $\Psi$ is first transformed according to Eq.~\eqref{psiV2}
\begin{equation}\label{psiV3}
\Psi = (r-r_h)^{-i\omega r_h}e^{i\omega(r-r_h)}r^{2i\omega r_h}\Phi  \nb
\end{equation}
then
\begin{equation}\label{RV3}
R=(1-z) \Phi ~. 
\end{equation}
The resulting master equation of $R$ as a function of $z$ is discretized into the matrix form Eq.~\eqref{masterMM}.

The boundary condition at $r=r_c$ can be determined by the ingoing wave in the tortoise coordinate
\begin{equation}\label{finV3}
f(r_*) = e^{-i\omega r_*}~,
\end{equation}
and one has
\bqn
\lb{Veff3bd}
\frac{R'}{R} = -1-2i (1+r_c) \omega  ~,
\eqn
where the derivative is taken with respect to $z$ at the point $z=0$.
This result is used to replace the first line of the matrix ${\cal M}$, and subsequently the quasinormal frequencies are evaluated.

To accertain the obtained results, one may follow the modified continued fraction method~\cite{agr-qnm-25, agr-qnm-star-08, agr-qnm-star-21}.
In this case, by carrying out a Taylor expansion of the wave function in $z$ around the cut $z=0$, the expansion coefficients are found to satisfy a four-term recurrence relation.
The latter can be further simplified by using the Gaussian elimination scheme to eliminate the ``off-diagonal'' elements associated with the fourth coefficient.
The standard continued fraction method can be readily applied to the resulting three-term recurrence relation in conjunction with the boundary condition Eq.~\eqref{Veff3bd}.

To be specific, by substituting the expasion of the wave function $\Psi$ into the master equation
\begin{equation}\label{psiV3b}
\Psi = (r-r_h)^{-i\omega r_h}e^{i\omega(r-r_h)}r^{2i\omega r_h}\sum_{n=0} a_n z^n ~,
\end{equation}
One finds the following recurence relations for $a_n$
\bqn\label{4trrgiusta}
\left\{ \begin{matrix}
\alpha_0 a_1+\beta_0 a_0=0& \\
\alpha_1 a_2+\beta_1 a_1+\gamma_1 a_0=0& \\
\alpha_n a_{n+1}+\beta_n a_n+\gamma_n a_{n-1}+\delta_n a_{n-2}=0 &  \ \ \ \mathrm{for}\ n\ge 2
\end{matrix} \right. \ ,
\eqn
where:
\bqn\label{primostep1}
&&\alpha_n=-n (1 + n) (-1 + r_c)\\
&&\beta_n=n (n (-3 + 2 r_c) + 2 (-2 + r_c^2) \rho) \nb\\
&&\gamma_n=-n^2 (-3 + r_c) + \ell r_c + \ell^2 r_c - \epsilon - 4 \rho + 4 \rho^2 +  4 r_c \rho^2 + n (-3 + r_c + 8 \rho) \nb\\
&&\delta_n=-n^2 + \epsilon + n (2 - 4 \rho) - 4 (-1 + \rho) \rho \nb ~,
\eqn
where $\rho=-i\omega$ and for axial gravitational perturbations $\epsilon=3$.

By employing the gaussian elimination scheme to eliminate $\delta_n$, one finds
\bqn\label{gausselim1}
&&\hat{\alpha}_0=\alpha_0,\qquad \hat{\beta}_0=\beta_0,\\
&&\hat{\alpha}_1=\alpha_1,\qquad \hat{\beta}_1=\beta_1,\qquad \hat{\gamma}_1=\gamma_1, \nb
\eqn
and for $n\geq 2$:
\bqn\label{gausselim2}
&&\hat{\alpha}_n=\alpha_n,\\
&&\hat{\beta}_n=\beta_n-\frac{\hat{\alpha}_{n-1}\delta_n}{\hat{\gamma}_{n-1}},\nb \\
&&\hat{\gamma}_n=\gamma_n-\frac{\hat{\beta}_{n-1}\delta_n}{\hat{\gamma}_{n-1}}, \nb\\
&&\hat{\delta}_n=0 .\nb 
\eqn
Eq.~\eqref{4trrgiusta} reduces to:
\bqn\lb{3trrgiusta} 
\left\{ \begin{matrix}
\hat\alpha_0 a_1+\hat\beta_0 a_0=0& \\
\hat{\alpha}_n a_{n+1}+\hat{\beta}_n a_n+\hat{\gamma}_n a_{n-1}=0 &  \ \ \ \mathrm{for}\ n\ge 1.
\end{matrix} \right. 
\eqn

Although Eq.~\eqref{3trrgiusta} is now identical to that of the original continued fraction method~\cite{agr-qnm-continued-fraction-01}, one should replace the first line of it with the counterpart of Eq.~\eqref{Veff3bd}.
To be specific, it can be shown that the boundary condition for $\Phi$ reads
\bqn
\lb{Veff3bd}
\frac{a_1}{a_0}=\frac{\Phi'}{\Phi} = -2 i (1+r_c) \omega  ~,
\eqn
for the above configuration.

\begin{table}[]
\caption{The low-lying QNM extracted for the truncated Regge-Wheeler potential Eq.~\eqref{Veff3} obtained by using the generalized matrix method (MM) and the modified continued fraction (MCF) method.}\lb{Tab3}
\begin{ruledtabular}
\renewcommand\arraystretch{2}
\begin{tabular}{cccccccc}
                        & $r_c$			& $3.0$		& 	          & $5.0$ 	&     		& $8.0$		     &   \\
                        & $n$			& $0$		& $1$	          & $0$ 	& $1$    		& $0$		     & $1$  \\
\hline
MM  	           & $\omega_{\mathrm{Re}}$ & $0.374316$   & $0.0936908 $  &  $0.235198 $  & $0.0337187 $ & $0.150689$  &  $0.0138962$      \\
                           & $\omega_{\mathrm{Im}}$ & $-0.210408 i$  & $-0.413573 i$ & $-0.143472 i$ & $-0.25736 i$ & $-0.0956327 i$ & $-0.16436 i$    \\
\hline
MCF	          & $\omega_{\mathrm{Re}}$ & $0.3743165$  & $0.09147263$  & $0.2351981 $  & $0.03345508 $ & $0.1506888$  & $0.01458402$    \\
                          & $\omega_{\mathrm{Im}}$ & $-0.2104078 i$ & $-0.40745718 i$ & $-0.1434721 i$ & $-0.2456574 i$  & $-0.09563271 i$ & $-0.153975 i$  
\end{tabular}
\end{ruledtabular}
\end{table} 


In Tab.~\ref{Tab3}, we show the obtained low-lying QNMs for different cuts $r_c$, which are compared against those obtained by the modified continued fraction method.
The results show reasonable agreement.
Since the algorithm of the modified continued fraction method is only convergent for $r_c > 2$, we have only chosen to present the results of the cases $r_c= 3, 5, 8$.

\begin{table}[]
\caption{The low-lying QNM extracted for the second effective potential Eq.~\eqref{Veff2} for $r_c=2.0$, obtained by using the matrix method with diffrent grid numbers.}\lb{Tab4}
\begin{ruledtabular}
\renewcommand\arraystretch{2}
\begin{tabular}{ccccc}
                        & $\mathrm{No.\ grids}$	&	$25$	&	$31$	          & $35$ 	   \\
\hline
n=0  	           & $\omega_{\mathrm{Re}}$ & $0.72595071360605285$   & $0.72595071361390585 $  &  $0.725950713613901981 $        \\
                           & $\omega_{\mathrm{Im}}$ & $-0.35517954712959586 i$  & $-0.35517954711878275 i$ & $-0.355179547118822025 i$     \\
\hline
n=1	          & $\omega_{\mathrm{Re}}$ & $0.81414693455219975$  & $0.81414693463923451$  & $0.814146934639889551 $      \\
                          & $\omega_{\mathrm{Im}}$ & $-1.10892664909983740 i$ & $-1.10892664951228270 i$ & $-1.108926649512453024 i$   \\
\hline
                        & $\mathrm{No.\ grids}$	&	$41$	&          $51$	 &  $81$   \\
\hline
n=0  	           & $\omega_{\mathrm{Re}}$ &  $0.72595071361390036 $ & $0.72595071361390017$  &  $0.72595071361390017$      \\
                           & $\omega_{\mathrm{Im}}$ &  $-0.35517954711882822 i$ & $-0.35517954711882856 i$ & $-0.35517954711882856 i$    \\
\hline
n=1	          & $\omega_{\mathrm{Re}}$ &  $0.81414693463990733 $ & $0.81414693463990736$  & $0.81414693463990736$    \\
                          & $\omega_{\mathrm{Im}}$ &  $-1.10892664951244659 i$  & $-1.10892664951244634 i$ & $-1.10892664951244634 i$  
\end{tabular}
\end{ruledtabular}
\end{table} 

It is also important to show the numerical convergence of the present scheme.
This is done by studying the dependence on the number of grids on which the expansions are performed.
The first two low-lying QNMs are evaluated for different numbers of grids and presented in Tab.~\ref{Tab4}.
Regarding the grid numbers, satisfactory convergence is observed.
As a comparison, we also apply the recurrence Taylor expansion scheme~\cite{agr-qnm-lq-07} up to the $120$th order and evaluate the two low-lying modes.
The resultling frequencies read $0.725950713615041178 - 0.35517954711870721 i$ and $0.81414693463986258 - 1.10892664951252702 i$.
By comparing against each other, the two approaches are manifestly consistent by eleven significant figures.

Before closing this section, we enumerate some of the favorable numerical features of the matrix method as a practical tool.
First, the matrix method involves inversions of matrices of significant size, which can be potentially time-consuming.
However, once the grids are given, such calculations can be performed and stored beforehand, independent of the specific form of metric.
As a result, the matrix method is rather efficient, and for the above examples, the calculations typically take a few seconds.
Second, as illustrated by the three examples above, the matrix method can typically be implemented by adopting a similar and essentially robust procedure for different cases.
In particular, if one replaces Eq.~\eqref{psiV2} by 
\begin{equation}\label{psiV2b}
\Psi = (r-r_h)^{-i\omega r_h}e^{i\omega(r-r_h)}r^{i\omega r_h}\Phi ,  \nb
\end{equation}
and accordingly substitutes Eq.~\eqref{Veff2bd} by
\bqn
\lb{Veff2bdc}
\frac{R'}{R} = 1+i (r_h+r_c) \omega  ~, \nb
\eqn
which still respect the ingoing wave boundary condition at the horizon, the matrix method yields essentially identical results.
However, this usually will not be the case for the continued fraction method, as a different wave function transform often gives rise to a different recurrence relation.
Lastly, we mention that the implementation of the orginal matrix method~\cite{agr-qnm-lq-matrix-01, agr-qnm-lq-matrix-02, agr-qnm-lq-matrix-03} can be found in the public repository of arXiv server.
Moreover, the source code used in the present paper has also been made available as supplemental material\footnote{The {\it Mathematica} notebook is published as a TeX-based supplemental material, accompanying the arXiv version of the paper.}.

\section{Further discussions and concluding remarks} \label{section5}

The present study is primarily motivated by the quasinormal modes and the associated structural instability in black hole effective potentials with discontinuity.
Aside from its mathematical simplification, the physical relevance of the discontinuity must also be justified.
In this regard, we argue that discontinuity is indeed a feature present in the metric of various feasible scenarios.
For exotic compact objects, discontinuous matter distribution is also introduced as one constructs the throat of traversable wormholes using the cut-and-paste procedure~\cite{agr-wormhole-10}.
Besides, discontinuity makes its appearance from a dynamic perspective.
For instance, in the framework of $\Lambda$CDM model, {\it cusp} was found in the halo profile~\cite{agr-dark-matter-06, agr-dark-matter-07}, which are largely compatible with the presence of dark halos~\cite{agr-dark-matter-08}.
In the outer region of the halo, a discontinuous ``splashback'' feature occurs due to the sudden drop in the density profile~\cite{agr-dark-matter-21, agr-dark-matter-24}, associated with the {\it caustic} formed by matter at their first apocenter after infall.
Also, in the study of the time evolution of a spherically collapsing matter~\cite{agr-collapse-thin-shell-11}, the interior metric is found to possess discontinuity.
Apart from the quasinormal mode structural instability, discontinuity in the metric is known for other interesting implications. 
It was shown that discontinuity in the effective potential furnishes a possible origin of black hole echoes~\cite{agr-qnm-echoes-20}, which was understood as due to the modification to the pole structure of Green's function.
The presence of a dusty thin shell was found to affect the black hole shadow non-trivially~\cite{agr-strong-lensing-shadow-35}. 

To summarize, in this study, we generalized the matrix method for black hole quasinormal modes to cope with discontinuous effective potential.
Mathematically, this is implemented by modifying the first and last line of the matrix in accordance with the boundary condition.
The obtained numerical results were compared with other approaches, which indicated that reasonable efficiency and precision could be achieved.
Nonetheless, similar to the continued fraction method, the present scheme is also somewhat constrained by the domain where the expansion is convergent.
Also, the high-order results presented in Tab.~\ref{Tab4} demand an adequate modification of the code to avoid the Runge phenomenon, which will be addressed in more detail elsewhere.
Recent results regarding the instability of the fundamental mode~\cite{agr-qnm-instability-15} invite further studies regarding different metrics and perturbations.
Since the presence of discontinuity poses a difficulty for direct applications of most approaches for quasinormal modes, the modified matrix method might serve as an alternative tool for relevant problems.

\section*{Acknowledgments}
This work is supported by the National Natural Science Foundation of China (NNSFC) under contract Nos. 11805166, 11925503, and 12175076.
We also gratefully acknowledge the financial support from
Funda\c{c}\~ao de Amparo \`a Pesquisa do Estado de S\~ao Paulo (FAPESP),
Funda\c{c}\~ao de Amparo \`a Pesquisa do Estado do Rio de Janeiro (FAPERJ),
Conselho Nacional de Desenvolvimento Cient\'{\i}fico e Tecnol\'ogico (CNPq),
Coordena\c{c}\~ao de Aperfei\c{c}oamento de Pessoal de N\'ivel Superior (CAPES),
A part of this work was developed under the project Institutos Nacionais de Ci\^{e}ncias e Tecnologia - F\'isica Nuclear e Aplica\c{c}\~{o}es (INCT/FNA) Proc. No. 464898/2014-5.
This research is also supported by the Center for Scientific Computing (NCC/GridUNESP) of the S\~ao Paulo State University (UNESP).

\bibliographystyle{h-physrev}
\bibliography{references_qian, references_more}

\begin{thebibliography}{10}

\bibitem{agr-LIGO-01}
Virgo, LIGO Scientific, B.~P. Abbott {\em et~al.},
\newblock Phys. Rev. Lett. {\bf 116}, 061102 (2016), arXiv:1602.03837.

\bibitem{agr-LIGO-02}
Virgo, LIGO Scientific, B.~P. Abbott {\em et~al.},
\newblock Phys. Rev. Lett. {\bf 116}, 221101 (2016), arXiv:1602.03841,
\newblock [Erratum: Phys. Rev. Lett.121,no.12,129902(2018)].

\bibitem{agr-LIGO-03}
Virgo, LIGO Scientific, B.~P. Abbott {\em et~al.},
\newblock Phys. Rev. Lett. {\bf 116}, 241103 (2016), arXiv:1606.04855.

\bibitem{agr-LIGO-04}
Virgo, LIGO Scientific, B.~P. Abbott {\em et~al.},
\newblock Phys. Rev. Lett. {\bf 119}, 141101 (2017), arXiv:1709.09660.

\bibitem{agr-LISA-01}
LISA, P.~Amaro-Seoane {\em et~al.},
\newblock (2017), arXiv:1702.00786.

\bibitem{agr-TianQin-01}
TianQin, J.~Luo {\em et~al.},
\newblock Class. Quant. Grav. {\bf 33}, 035010 (2016), arXiv:1512.02076.

\bibitem{agr-Taiji-01}
W.-R. Hu and Y.-L. Wu,
\newblock Natl. Sci. Rev. {\bf 4}, 685 (2017).

\bibitem{agr-qnm-review-02}
H.-P. Nollert,
\newblock Class. Quant. Grav. {\bf 16}, R159 (1999).

\bibitem{agr-qnm-review-03}
E.~Berti, V.~Cardoso, and A.~O. Starinets,
\newblock Class. Quant. Grav. {\bf 26}, 163001 (2009), arXiv:0905.2975.

\bibitem{agr-qnm-review-06}
B.~Wang,
\newblock Braz. J. Phys. {\bf 35}, 1029 (2005), arXiv:gr-qc/0511133.

\bibitem{agr-bh-nohair-01}
J.~D. Bekenstein,
\newblock Phys. Rev. Lett. {\bf 28}, 452 (1972).

\bibitem{agr-bh-nohair-04}
J.~D. Bekenstein,
\newblock Phys. Rev. {\bf D51}, R6608 (1995).

\bibitem{Santos:2019yzk}
E.~C. Santos, J.~C. Fabris, and J.~A. de~Freitas~Pacheco,
\newblock (2019), arXiv:1903.04874.

\bibitem{Campos:2021sff}
J.~A.~V. Campos, M.~A. Anacleto, F.~A. Brito, and E.~Passos,
\newblock Sci. Rep. {\bf 12}, 8516 (2022), arXiv:2103.10659.

\bibitem{Anacleto:2021qoe}
M.~A. Anacleto, J.~A.~V. Campos, F.~A. Brito, and E.~Passos,
\newblock Annals Phys. {\bf 434}, 168662 (2021), arXiv:2108.04998.

\bibitem{Rincon:2018sgd}
A.~Rinc\'on and G.~Panotopoulos,
\newblock Phys. Rev. D {\bf 97}, 024027 (2018), arXiv:1801.03248.

\bibitem{Panotopoulos:2017hns}
G.~Panotopoulos and A.~Rinc\'on,
\newblock Int. J. Mod. Phys. D {\bf 27}, 1850034 (2017), arXiv:1711.04146.

\bibitem{Rincon:2018ktz}
A.~Rinc\'on and G.~Panotopoulos,
\newblock Eur. Phys. J. C {\bf 78}, 858 (2018), arXiv:1810.08822.

\bibitem{Ponglertsakul:2018smo}
S.~Ponglertsakul, P.~Burikham, and L.~Tannukij,
\newblock Eur. Phys. J. C {\bf 78}, 584 (2018), arXiv:1803.09078.

\bibitem{Fernando:2022wlm}
S.~Fernando, P.~A. Gonz\'alez, and Y.~V\'asquez,
\newblock Eur. Phys. J. C {\bf 82}, 600 (2022), arXiv:2204.02755.

\bibitem{Gonzalez:2022upu}
P.~A. Gonz\'alez, E.~Papantonopoulos, J.~Saavedra, and Y.~V\'asquez,
\newblock JHEP {\bf 06}, 150 (2022), arXiv:2204.01570.

\bibitem{Rincon:2021gwd}
A.~Rincon, P.~A. Gonzalez, G.~Panotopoulos, J.~Saavedra, and Y.~Vasquez,
\newblock (2021), arXiv:2112.04793.

\bibitem{agr-TDI-review-01}
M.~Tinto and S.~V. Dhurandhar,
\newblock Living Rev. Rel. {\bf 17}, 6 (2014).

\bibitem{agr-TDI-review-02}
M.~Tinto and S.~V. Dhurandhar,
\newblock Living Rev. Rel. {\bf 24}, 1 (2021).

\bibitem{agr-TDI-Wang-01}
P.-P. Wang, Y.-J. Tan, W.-L. Qian, and C.-G. Shao,
\newblock Phys. Rev. D {\bf 104}, 082002 (2021), arXiv:2106.02236.

\bibitem{agr-SNR-Wang-01}
P.-P. Wang, Y.-J. Tan, W.-L. Qian, and C.-G. Shao,
\newblock Phys. Rev. D {\bf 103}, 063021 (2021).

\bibitem{agr-SNR-Wang-02}
P.-P. Wang, Y.-J. Tan, W.-L. Qian, and C.-G. Shao,
\newblock Phys. Rev. D {\bf 104}, 023002 (2021).

\bibitem{agr-SNR-05}
O.~Dreyer {\em et~al.},
\newblock Class. Quant. Grav. {\bf 21}, 787 (2004), arXiv:gr-qc/0309007.

\bibitem{agr-SNR-06}
E.~Berti, V.~Cardoso, and C.~M. Will,
\newblock Phys. Rev. {\bf D73}, 064030 (2006), arXiv:gr-qc/0512160.

\bibitem{agr-SNR-10}
M.~Giesler, M.~Isi, M.~A. Scheel, and S.~Teukolsky,
\newblock Phys. Rev. X {\bf 9}, 041060 (2019), arXiv:1903.08284.

\bibitem{agr-SNR-18}
M.~Cabero {\em et~al.},
\newblock Phys. Rev. D {\bf 101}, 064044 (2020), arXiv:1911.01361.

\bibitem{agr-SNR-20}
A.~Dhani,
\newblock Phys. Rev. D {\bf 103}, 104048 (2021), arXiv:2010.08602.

\bibitem{agr-SNR-36}
H.~Liu, C.~Zhang, Y.~Gong, B.~Wang, and A.~Wang,
\newblock Phys. Rev. {\bf D102}, 124011 (2020), arXiv:2002.06360.

\bibitem{agr-bh-thermodynamics-12}
M.~Visser,
\newblock Phys. Rev. {\bf D46}, 2445 (1992), arXiv:hep-th/9203057.

\bibitem{agr-qnm-33}
P.~T. Leung, Y.~T. Liu, W.~M. Suen, C.~Y. Tam, and K.~Young,
\newblock Phys. Rev. Lett. {\bf 78}, 2894 (1997), arXiv:gr-qc/9903031.

\bibitem{agr-qnm-34}
P.~T. Leung, Y.~T. Liu, W.~M. Suen, C.~Y. Tam, and K.~Young,
\newblock Phys. Rev. {\bf D59}, 044034 (1999), arXiv:gr-qc/9903032.

\bibitem{agr-qnm-54}
E.~Barausse, V.~Cardoso, and P.~Pani,
\newblock Phys. Rev. {\bf D89}, 104059 (2014), arXiv:1404.7149.

\bibitem{agr-qnm-35}
H.-P. Nollert,
\newblock Phys. Rev. {\bf D53}, 4397 (1996), arXiv:gr-qc/9602032.

\bibitem{agr-qnm-36}
H.-P. Nollert and R.~H. Price,
\newblock J. Math. Phys. {\bf 40}, 980 (1999), arXiv:gr-qc/9810074.

\bibitem{agr-qnm-50}
R.~G. Daghigh, M.~D. Green, and J.~C. Morey,
\newblock Phys. Rev. {\bf D101}, 104009 (2020), arXiv:2002.07251.

\bibitem{agr-qnm-lq-03}
W.-L. Qian, K.~Lin, C.-Y. Shao, B.~Wang, and R.-H. Yue,
\newblock Phys. Rev. {\bf D103}, 024019 (2021), arXiv:2009.11627.

\bibitem{agr-qnm-continued-fraction-02}
H.-P. Nollert,
\newblock Phys. Rev. {\bf D47}, 5253 (1993).

\bibitem{agr-qnm-continued-fraction-03}
L.~Motl,
\newblock Adv. Theor. Math. Phys. {\bf 6}, 1135 (2003), arXiv:gr-qc/0212096.

\bibitem{agr-qnm-instability-07}
J.~L. Jaramillo, R.~Panosso~Macedo, and L.~Al~Sheikh,
\newblock Phys. Rev. X {\bf 11}, 031003 (2021), arXiv:2004.06434.

\bibitem{agr-qnm-instability-13}
J.~L. Jaramillo, R.~Panosso~Macedo, and L.~A. Sheikh,
\newblock (2021), arXiv:2105.03451.

\bibitem{agr-qnm-instability-14}
K.~Destounis, R.~P. Macedo, E.~Berti, V.~Cardoso, and J.~L. Jaramillo,
\newblock Phys. Rev. D {\bf 104}, 084091 (2021), arXiv:2107.09673.

\bibitem{agr-qnm-hyperboloidal-01}
A.~Zenginoğlu,
\newblock Phys. Rev. {\bf D83}, 127502 (2011), arXiv:1102.2451.

\bibitem{agr-qnm-instability-15}
M.~H.-Y. Cheung, K.~Destounis, R.~P. Macedo, E.~Berti, and V.~Cardoso,
\newblock Phys. Rev. Lett. {\bf 128}, 111103 (2022), arXiv:2111.05415.

\bibitem{agr-qnm-WKB-01}
B.~F. Schutz and C.~M. Will,
\newblock Astrophys. J. {\bf 291}, L33 (1985).

\bibitem{agr-qnm-WKB-02}
S.~Iyer and C.~M. Will,
\newblock Phys. Rev. {\bf D35}, 3621 (1987).

\bibitem{agr-qnm-WKB-03}
R.~A. Konoplya,
\newblock Phys. Rev. {\bf D68}, 024018 (2003), arXiv:gr-qc/0303052.

\bibitem{agr-qnm-WKB-05}
J.~Matyjasek and M.~Telecka,
\newblock Phys. Rev. {\bf 100}, 124006 (2019), arXiv:1908.09389.

\bibitem{agr-qnm-40}
L.~Motl and A.~Neitzke,
\newblock Adv. Theor. Math. Phys. {\bf 7}, 307 (2003), arXiv:hep-th/0301173.

\bibitem{agr-qnm-star-07}
K.~D. Kokkotas and B.~F. Schutz,
\newblock Mon. Not. Roy. Astron. Soc. {\bf 255}, 119 (1992).

\bibitem{agr-qnm-star-08}
M.~Leins, H.~P. Nollert, and M.~H. Soffel,
\newblock Phys. Rev. {\bf D48}, 3467 (1993).

\bibitem{agr-qnm-continued-fraction-01}
E.~W. Leaver,
\newblock Proc. Roy. Soc. Lond. {\bf A402}, 285 (1985).

\bibitem{agr-qnm-star-25}
P.~Pani, E.~Berti, V.~Cardoso, Y.~Chen, and R.~Norte,
\newblock Phys. Rev. {\bf D80}, 124047 (2009), arXiv:0909.0287.

\bibitem{agr-qnm-lq-matrix-01}
K.~Lin and W.-L. Qian,
\newblock (2016), arXiv:1609.05948.

\bibitem{agr-qnm-lq-matrix-02}
K.~Lin and W.-L. Qian,
\newblock Class. Quant. Grav. {\bf 34}, 095004 (2017), arXiv:1610.08135.

\bibitem{agr-qnm-lq-matrix-03}
K.~Lin, W.-L. Qian, A.~B. Pavan, and E.~Abdalla,
\newblock Mod. Phys. Lett. {\bf A32}, 1750134 (2017), arXiv:1703.06439.

\bibitem{agr-qnm-lq-matrix-04}
K.~Lin and W.-L. Qian,
\newblock Chin. Phys. {\bf C43}, 035105 (2019), arXiv:1902.08352.

\bibitem{agr-collapse-thin-shell-03}
W.~Israel,
\newblock Nuovo Cim. B {\bf 44S10}, 1 (1966),
\newblock [Erratum: Nuovo Cim.B 48, 463 (1967)].

\bibitem{book-general-relativity-Poisson}
E.~Poisson,
\newblock {\em {A Relativist's Toolkit: The Mathematics of Black-Hole
  Mechanics}} (Cambridge University Press, 2009).

\bibitem{agr-qnm-Poschl-Teller-02}
V.~Ferrari and B.~Mashhoon,
\newblock Phys. Rev. {\bf D30}, 295 (1984).

\bibitem{agr-qnm-lq-07}
K.~Lin and W.-L. Qian,
\newblock in preparation .

\bibitem{agr-qnm-25}
E.~W. Leaver,
\newblock Phys. Rev. {\bf D41}, 2986 (1990).

\bibitem{agr-qnm-star-21}
O.~Benhar, E.~Berti, and V.~Ferrari,
\newblock Mon. Not. Roy. Astron. Soc. {\bf 310}, 797 (1999),
  arXiv:gr-qc/9901037.

\bibitem{agr-wormhole-10}
M.~Visser,
\newblock Phys. Rev. {\bf D39}, 3182 (1989), arXiv:0809.0907.

\bibitem{agr-dark-matter-06}
J.~F. Navarro, C.~S. Frenk, and S.~D.~M. White,
\newblock Astrophys. J. {\bf 462}, 563 (1996), arXiv:astro-ph/9508025.

\bibitem{agr-dark-matter-07}
B.~Moore, F.~Governato, T.~R. Quinn, J.~Stadel, and G.~Lake,
\newblock Astrophys. J. Lett. {\bf 499}, L5 (1998), arXiv:astro-ph/9709051.

\bibitem{agr-dark-matter-08}
O.~Valenzuela {\em et~al.},
\newblock Astrophys. J. {\bf 657}, 773 (2007), arXiv:astro-ph/0509644.

\bibitem{agr-dark-matter-21}
B.~Diemer and A.~V. Kravtsov,
\newblock Astrophys. J. {\bf 789}, 1 (2014), arXiv:1401.1216.

\bibitem{agr-dark-matter-24}
S.~Adhikari, N.~Dalal, and R.~T. Chamberlain,
\newblock JCAP {\bf 1411}, 019 (2014), arXiv:1409.4482.

\bibitem{agr-collapse-thin-shell-11}
H.~Kawai, Y.~Matsuo, and Y.~Yokokura,
\newblock Int. J. Mod. Phys. {\bf A28}, 1350050 (2013), arXiv:1302.4733.

\bibitem{agr-qnm-echoes-20}
H.~Liu {\em et~al.},
\newblock Phys. Rev. {\bf D104}, 044012 (2021), arXiv:2104.11912.

\bibitem{agr-strong-lensing-shadow-35}
W.-L. Qian, S.~Chen, C.-G. Shao, B.~Wang, and R.-H. Yue,
\newblock Eur. Phys. J. C {\bf 82}, 91 (2022), arXiv:2102.03820.

\end{thebibliography}

\end{document}